\documentclass[aps,eqsecnum,nofootinbib,prd,showpacs]{revtex4}
\usepackage{amsmath}
\usepackage{amssymb}
\usepackage{amsfonts}
\usepackage{tensor}
\usepackage{graphicx}
\usepackage{appendix}
\usepackage[pdftex,
  pdftitle={Gravitational Interactions in a General Multibrane Model},
  pdfauthor={Jolyon K Bloomfield, Eanna E Flanagan},
  bookmarks,bookmarksopen=false,
  pdfstartview={FitH}
]{hyperref}

\numberwithin{equation}{section}
\newcommand{\n}[2][{}]{\ensuremath{\tensor*[^{n}]{#2}{#1}}}
\newcommand{\np}[2][{}]{\ensuremath{\tensor*[^{n+1}]{#2}{#1}}}

\newcommand{\lrn}[1]{\ensuremath{\tensor*{#1}{_n}}}
\newcommand{\lrnm}[1]{\ensuremath{\tensor*{#1}{_{n-1}}}}
\newcommand{\lrnp}[1]{\ensuremath{\tensor*{#1}{_{n+1}}}}

\newcommand{\kapfs}{\ensuremath{\tensor*{\kappa}{_5^2}}}
\newcommand{\bn}{{\ensuremath{{{\cal{B}}_{n}}}}}
\newcommand{\rn}{{\ensuremath{{{\cal{R}}_{n}}}}}
\newcommand{\E}{\mu}

\begin{document}

\title{Gravitational Interactions in a General Multibrane Model}

\author{Jolyon K. Bloomfield}
\email{jkb84@cornell.edu}
\affiliation{Physics Department, Cornell University, Ithaca, New York 14853, USA}

\author{\'Eanna \'E. Flanagan}
\email{flanagan@astro.cornell.edu}
\affiliation{Center for Radiophysics and Space Research, Cornell University, Ithaca, New York 14853, USA}
\affiliation{Newman Laboratory for Elementary Particle Physics, Cornell University, Ithaca, New York 14853, USA}

\date{September 23rd, 2010}

\begin{abstract}
The gravitational interactions of the four-dimensional effective theory describing a general $N$-brane model in five dimensions without radion stabilization are analyzed. Both uncompactified and orbifolded models are considered. The parameter space is constrained by requiring that there be no ghost modes in the theory, and that the Eddington parameterized post-Newtonian parameter $\gamma$ be consistent with observations. We show that we must reside on the brane on which the warp factor is maximized. The resultant theory contains $N-1$ radion modes in a nonlinear sigma model, with the target space being a subset of hyperbolic space. Imposing observational constraints on the relative strengths of gravitational interactions of dark and visible matter shows that at least 99.8\% of the dark matter must live on our brane in this model.
\end{abstract}


\maketitle

\renewcommand\thesection{\arabic{section}}
\renewcommand\thesubsection{\Alph{subsection}}
\renewcommand\thesubsubsection{\Roman{subsubsection}}

\section{Introduction and Summary}

Over the past ten years, there has been a large amount of interest in extra-dimensional models of the Universe. Models such as the ADD model \cite{ADD1998, Antoniadis1998} and the Randall-Sundrum model \cite{Randall1999, Randall1999a} have demonstrated the possibility of a solution to the hierarchy problem through a modification of the fundamental gravitational scale, and the potential to provide interesting models of dark matter.

Building on the success of the Randall-Sundrum model, many papers have considered various extensions to it, including bulk fields \cite{Goldberger1999}, radion stabilization mechanisms \cite{Goldberger1999a, Goldberger2000}, and models including more than one or two branes \cite{Zen2005, Cotta-Ramusino2004, Kogan2000, Kogan2001, Kogan2002, Flanagan2001a}. A wealth of knowledge of the phenomenology of these models has been accumulated \cite{Brax2003, Csaki2002, Maartens2004, Perez-Lorenzana2004}. A variety of techniques to analyze such models have been devised, such as linearized analyses \cite{Garriga2000, Kudoh2001, Kudoh2002, Carena2005, Callin2005}, the ``covariant curvature method'' \cite{Shiromizu2000}, the ``gradient expansion method'' \cite{Kanno2002a, Kanno2002, Soda2003}, and derivative expansion methods \cite{Wiseman2002}. However, relatively little work has gone into analyzing a general $N$-brane model, and of the techniques mentioned previously, only linearized analyses are well equipped to analyze such a situation. Even then, linearized analyses require a background solution to perturb.

In a previous paper \cite{Bloomfield2010}, we proposed an approximation scheme based upon a two-lengthscale expansion which can be used to evaluate a four-dimensional low energy action for five-dimensional braneworld models, and demonstrated its application to an uncompactified $N$-brane model. In this paper, we analyze the physics of the model, starting from the four-dimensional action previously found, and also generalize the results to orbifolded models. Our motivation in analyzing general $N$-brane models is to determine whether the presence of extra branes may overcome some of the constraints the RS-I and RS-II models have, particularly with regards to radion stabilization requirements. We investigate the parameter space of the general model, and find regions in which the theory has no ghosts. The parameter space is further refined by imposing observational constraints from Solar System tests of gravity. We consider the possibility of placing dark matter and Standard Model fields on separate branes, and by comparing to observational data, find that the vast majority of the dark matter must reside on our brane in the models considered.

This paper is organized as follows. We begin in Section \ref{sec:recap} by describing the model and recalling the results from \cite{Bloomfield2010} upon which we build in this paper. In Section \ref{sec:fieldspace}, we diagonalize the field space metric of the radion modes for the entire parameter space. Only certain subsets of the model parameter space give rise to ghost-free four-dimensional theories; we derive the corresponding conditions in Section \ref{sec:reasonablesection}. In Section \ref{sec:PhysAction}, we derive the physical four-dimensional low-energy action. Finally, in Section \ref{sec:coupling}, we determine the observational consequences of this action, calculating the Eddington parameterized post-Newtonian parameter $\gamma$ and the Newton's matrix for gravitational interactions between branes.

The methodology considered here is also applicable to orbifolded models. In Appendix \ref{app:orbifold}, we show that the low energy theory for orbifolded models is very similar to that for uncompactified models. In Appendix \ref{app:KKmodes}, we discuss the spectrum of Kaluza-Klein modes in both orbifolded and uncompactified multibrane models.

\section{The Four-Dimensional Low Energy Action\label{sec:recap}}
In this section, we recall the results from \cite{Bloomfield2010} that we use in the rest of this paper.

Consider a five-dimensional model containing $N$ four-dimensional branes, each with their own brane tension $\sigma_n$. We denote the $n^{\mathrm{th}}$ brane by \bn, where $n$ is an integer with $0 \le n \le N-1$. In between each brane there exists a bulk region of spacetime, which we denote ${\cal{R}}_0, \ldots , {\cal{R}}_N$, with \rn\ lying between branes $n-1$ and $n$. In each bulk region \rn\ we allow for a bulk cosmological constant \lrn{\Lambda}. We allow matter fields represented by $\n\phi$ to reside on each brane, and include corresponding general four-dimensional matter terms in the action. The general five-dimensional action we begin with is
\begin{align}
S \left[ \tensor{g}{_{\Gamma \Sigma}}, \tensor[^n]{x}{^\Gamma}, \n\phi \right] ={}& \int d^5 x \sqrt{-g} \left(\frac{R^{(5)}[\tensor{g}{_{\Gamma \Sigma}}]}{2 \kapfs} - \Lambda(x^\Gamma) \right) - \sum_{n = 0}^{N-1} \lrn{\sigma} \int d^4 \lrn{w} \sqrt{-\n{h}} + \sum_{n = 0}^{N-1} \n[_m]{S}[\n[_{ab}]{h}, \n{\phi}],
\end{align}
where $g_{\Gamma \Sigma}$ is the bulk metric, $\Gamma, \Sigma$ run from 0 to 4, $\n[^\Gamma]{x} = \n[^\Gamma]{x} (\tensor*{w}{_n^a})$ gives the location of the $n^{\mathrm{th}}$ brane \bn\ in terms of the brane coordinates $\tensor*{w}{_n^a}$ with $0 \le \alpha \le 3$, \kapfs\ is the five-dimensional Newton's constant, $\n[_{ab}]{h}$ is the induced metric on a brane, and $\Lambda(x^\Gamma)$ takes the value $\Lambda_n$ in the appropriate regions.

\begin{figure}[t]
    \centering
        \includegraphics[width = 0.7\textwidth]{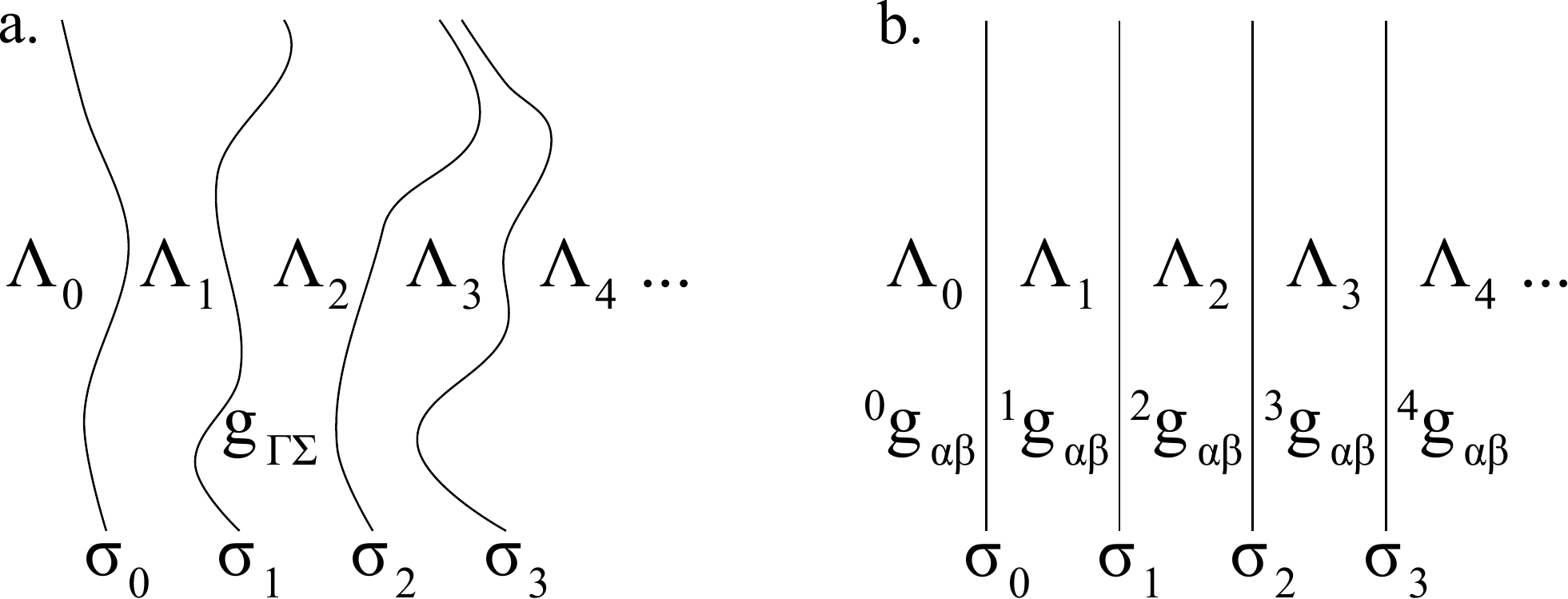}
    \caption{An illustration of the model a) before and b) after gauge fixing. The bulk cosmological constants \lrn{\Lambda}, brane tensions \lrn{\sigma}, and metrics \n[_{\alpha \beta}]{g} are labeled.} \label{fig:setup}
\end{figure}

Our first step is to partially fix the gauge so that the branes lie at $y=0, 1, \ldots, N-1$, as seen in Fig. \ref{fig:setup} \cite{Bloomfield2010}. Next, applying a separation of lengthscales technique, we find the general bulk metric which solves the high energy dynamics of the model,
\begin{align}
\n[^2]{ds} = e^{\chi(x^c, y)} \hat{\gamma}_{ab} (x^c) dx^a dx^b + \frac{\chi^2_{,y} (x^c, y)}{4 k_n^2} dy^2. \label{eq:initialmetric}
\end{align}
This metric is defined in each individual bulk region. The constants $k_n$ are defined by
\begin{align}
k_n = \sqrt{\frac{-\kapfs \lrn \Lambda}{6}},
\end{align}
and the function $\chi(x^c,y)$ is continuous across all branes, although it has discontinuities in its first derivative in $y$. The solution requires that the brane tensions are tuned to some fixed values,
\begin{equation}
\lrn{k} \lrn{P} - \lrnp{k} \lrnp{P} = \frac{1}{3} \kapfs \lrn \sigma, \label{eq:branetunings}
\end{equation}
where $P_n = \mathrm{sgn} (\chi_{,y})$ in \rn\ (where $P_n$ is constant in each bulk region).

Having solved for the high energy dynamics, we now proceed to integrate them out of the theory, thereby generating a four-dimensional low energy action. We define the fixed parameters
\begin{align}
A_n ={}& \left| \frac{1}{k_n P_n} - \frac{1}{k_{n+1} P_{n+1}} \right|, \label{eq:def:A}
\\
\epsilon_n ={}& \mathrm{sgn} \left( \frac{1}{k_n P_n} - \frac{1}{k_{n+1} P_{n+1}} \right), \label{eq:def:epsilon}
\end{align}
for $0 \le n \le N-1$. It is useful to note that $\epsilon_n = - \mathrm{sgn} \left( \sigma_n P_n P_{n+1} \right)$ by Eq. \eqref{eq:branetunings}. The values of the function $\chi(x^c, n)$ evaluated on the branes become $N$ scalar fields in the four-dimensional action, and we denote these by
\begin{align}
\Psi_n ={}& \sqrt{A_n e^{\chi_n}},
\end{align}
where we use $\chi_n = \chi(x^a, n)$. There is residual parameterization freedom which implies that one of the fields $\Psi_n$ is nondynamical, but before we fix this freedom, we first give the four-dimensional low-energy action using the definitions so far. It is given by
\begin{align}
 S \left[\hat{\gamma}_{ab}, \lrn \Psi, \n\phi\right] = \int d^4 x \sqrt{-\hat{\gamma}} \frac{1}{4 \kappa_5^2} \left[ R^{(4)} \left[ \hat{\gamma}_{ab} \right] \left( \sum_{n = 0}^{N-1} \epsilon_n \Psi_n^2 \right) + 6 \sum_{n = 0}^{N-1} \epsilon_n (\hat{\nabla}^a \Psi_n) (\hat{\nabla}_a \Psi_n) \right] + \sum_{n = 0}^{N-1} \n[_m]S \left[ \frac{\Psi_n^2}{A_n} \hat{\gamma}_{ab}, \n\phi \right] \label{eq:4DJordanBeforeGauge}
\end{align}
where indices are raised and lowered using the four-dimensional metric $\hat{\gamma}_{ab}$, and $\hat{\nabla}_a$ is the covariant derivative associated with this same metric. In Appendix \ref{app:orbifold}, we show that an orbifolded $N$-brane model gives rise to this same four-dimensional low-energy action with a rescaling of some parameters. Most of what follows from here is the same for orbifolded and uncompactified models.

The residual parameterization freedom is
\begin{align}
\chi(x^a, y) \rightarrow {}& \chi(x^a, y) + \delta \chi (x^a)
\\
\hat{\gamma}_{ab}(x^a) \rightarrow {}& \hat{\gamma}_{ab} e^{- \delta \chi (x^a)},
\end{align}
under which the metric \eqref{eq:initialmetric} is invariant. We can fix this freedom by specifying the value of $\chi(x^a, n)$ for any $n$. In order to remain general, let us choose $\chi(x^a, T) = 0$, for some $T$ with $0 \le T \le N-1$. This causes the field $\Psi_T$ to become non-dynamical.

Some further field redefinitions now simplify the action. Let
\begin{align}
B_n ={}& \frac{A_n}{A_T}, \label{eq:def:B}
\\
\psi_n ={}& \sqrt{B_n e^{\chi_n}} = \frac{\Psi_n}{\sqrt{A_T}} \label{eq:def:psi}.
\end{align}
Our dynamical scalar fields are now $\psi_n$, $0 \le n \le N - 1, n \neq T$. Finally, we can define a four-dimensional effective Newton's constant as
\begin{equation}
\frac{1}{2 \kappa_4^2} = \frac{1}{4 \kappa_5^2} A_T. \label{eq:def:4dG}
\end{equation}
The action with these definitions is
\begin{align}
 S \left[\hat{\gamma}_{ab}, \lrn \psi, \n\phi\right] ={}& \int d^4 x \sqrt{-\hat{\gamma}} \frac{\epsilon_T}{2 \kappa_4^2} \left[ R^{(4)} \left[ \hat{\gamma}_{ab} \right] \left( 1 + \sum_{\substack{n = 0 \\ n \neq T}}^{N-1} \epsilon_T \epsilon_n \psi_n^2 \right) + 6 \sum_{n = 1}^{N-1} \epsilon_T \epsilon_n (\hat{\nabla}^a \psi_n) (\hat{\nabla}_a \psi_n) \right] \nonumber
\\
 & {} + \tensor[^T]S{_m}[\tensor{\hat{\gamma}}{_{ab}}, \tensor[^T]{\phi}{}] + \sum_{\substack{n = 0 \\ n \neq T}}^{N-1} \n[_m]S \left[ \frac{\psi_n^2}{B_n} \hat{\gamma}_{ab}, \n\phi \right]. \label{eq:4DJordanOriginal}
\end{align}

We next discuss the transformation to the Einstein conformal frame. Let $P$ be the number of elements of the set $\{\epsilon_T \epsilon_n, 0 \le n \le N-1, n \neq T \}$ for which $\epsilon_T \epsilon_n = +1$, corresponding to the number of scalar fields with positive coefficients in the action. Note that $0 \le P \le N-1$. Also, let $M = N-1-P$ be the number of elements with $\epsilon_T \epsilon_n$ negative, corresponding to the number of scalar fields with negative coefficients. It is convenient to relabel the fields $\{ \psi_n \}$ based on which have positive kinetic coefficient $(\psi_1, \ldots, \psi_P)$ and which have negative kinetic coefficient $(\psi_{P+1}, \ldots, \psi_{P+M})$, based on the action \eqref{eq:4DJordanOriginal}. We now define new coordinates $\zeta, \theta_1, \ldots, \theta_{P-1}$ and $\eta, \lambda_1, \ldots, \lambda_{M-1}$, such that
\begin{subequations}
\label{eq:def:scalarfields}
\begin{align}
(\psi_1, \ldots, \psi_P) = {}& \zeta \left(\cos(\theta_1), \sin(\theta_1) \cos(\theta_2), \ldots, \sin(\theta_1) \sin(\theta_2) \cdots \sin(\theta_{P-1})\right),
\\
(\psi_{P+1}, \ldots, \psi_{P+M}) = {}& \eta \left(\cos(\lambda_1), \sin(\lambda_1) \cos(\lambda_2), \ldots, \sin(\lambda_1) \sin(\lambda_2) \cdots \sin(\lambda_{M-1})\right). \label{eq:def:scalarfieldsm}
\end{align}
\end{subequations}
We choose $\eta$, $\zeta > 0$. All of the angular fields ($\theta_i$ and $\lambda_j$) have a domain of $0$ to $\pi/2$, as each $\psi_n$ is positive. We define the function
\begin{align}
\Theta = 1 + \sum_{\substack{n = 0 \\ n \neq T}}^{N-1} \epsilon_T \epsilon_n \psi_n^2 = 1 + \zeta^2 - \eta^2, \label{eq:def:theta}
\end{align}
and transform to the Einstein conformal frame using the conformal transformation $g_{ab} = \hat{\gamma}_{ab} |\Theta|$.

Using these field definitions, the four-dimensional low energy action can be written in Einstein conformal frame as
\begin{align}
 S[g_{ab}, \Phi^A, \n\phi] ={}& \int d^4 x \sqrt{-g} \epsilon_T \mathrm{sgn}\left(\Theta\right)\left[ \frac{R^{(4)} [g_{ab}]}{2 \kappa_4^2} - \frac{1}{2} \gamma_{AB}(\Phi^C) g^{ab} \nabla_a \Phi^A \nabla_b \Phi^B \right] + \sum_{n=0}^{N-1} \n[_m]S \left[ e^{2 \alpha_n (\Phi^C)} g_{ab}, \n\phi \right]. \label{eq:actionEinstein}
\end{align}
Here, $R^{(4)}$ and $\nabla_a$ are associated with the metric $g_{ab}$, $\Phi^A \equiv \left( \zeta, \eta, \theta_1, \ldots, \theta_{P-1}, \lambda_1, \ldots, \lambda_{M-1}\right) $, and $\gamma_{AB} (\Phi^C)$ is the metric on field space, given by
\begin{align}
d\sigma^2 = \gamma_{AB} d\Phi^A d\Phi^B = \frac{\E^2}{\Theta} \left[ - d\zeta^2 \left( \frac{1 - \eta^2}{\Theta} \right) - \zeta^2 d\Omega_p^2 + d\eta^2 \left( \frac{1 + \zeta^2}{\Theta} \right) + \eta^2 d\Omega_m^2 - \frac{2 \eta \zeta}{\Theta} d\eta d\zeta \right], \label{eq:FullFieldMetric}
\end{align}
where $d\Omega_p^2 = d\theta_1^2 + \sin^2(\theta_1) d\theta_2^2 + \ldots$ is the metric on the unit $(P-1)$-sphere, and similarly for $d\Omega_m^2$. The parameter $\mu$ is defined by $\mu = \sqrt{6}/\kappa_4$. The coupling functions $\alpha_n(\Phi^C)$ are given by
\begin{subequations}
\begin{align}
e^{2 \alpha_T} = {}& \frac{1}{|\Theta|},
\\
e^{2 \alpha_n} = {}& \frac{1}{|\Theta|} \frac{\psi_n^2}{B_n}, \ 0 \le n \le N-1, n \neq T,
\end{align}
\end{subequations}
where $B_n$ is given by Eq. \eqref{eq:def:B}, and $\psi_n (\Phi^C)$ is defined by the relevant expression in Eq. \eqref{eq:def:scalarfields}. 

\section{Parameterization of Field Space\label{sec:fieldspace}}
In this section, we find coordinates on field space which diagonalize the field space metric \eqref{eq:FullFieldMetric}. We look at two special cases before analyzing the general case.

\subsection{Negative Definite Field Space Metric}
In the case $M=0$, the general metric reduces to
\begin{align}
\frac{(1 + \zeta^2)}{\E^2} d\sigma^2 ={}& - \frac{1}{1 + \zeta^2} d\zeta^2 - \zeta^2 d\Omega_p^2.
\end{align}
This can be rewritten as
\begin{align}
d\sigma^2 ={}& - da^2 - \E^2 \sin^2\left(\frac{a}{\E}\right) d\Omega_p^2, \label{eq:MetMZero}
\end{align}
where $a = \E \tan^{-1}(\zeta)$, with $0 \le a \le \pi \E/2$.


\subsection{Positive Definite Field Space Metric\label{sec:PisZero}}
In the case of $P=0$, the general metric reduces to
\begin{align}
\frac{(1 - \eta^2)}{\E^2} d\sigma^2 ={}& d\eta^2 \frac{1}{1 - \eta^2} + \eta^2 d\Omega_n^2. \label{eq:PisZeroMetric}
\end{align}

For the case where $\eta < 1$, this can be rewritten as
\begin{align}
d\sigma^2 ={}& da^2 + \E^2 \sinh^2\left(\frac{a}{\E}\right) d\Omega_n^2, \label{eq:MetPZeroMinus}
\end{align}
where $a = \E \tanh^{-1} (\eta)$, with $0 < a < \infty$. This is shown in Section \ref{sec:reasonablesection} to be the only physically relevant case.


For the case of $\eta > 1$, the metric \eqref{eq:FullFieldMetric} can be rewritten as
\begin{align}
d\sigma^2 ={}& da^2 - \E^2 \cosh^2\left(\frac{a}{\E}\right) d\Omega_n^2, \label{eq:MetPZeroMixed}
\end{align}
where $a = \E \coth^{-1} (\eta)$, and $0 < a < \infty$.


We see that the two cases $\eta >1$ and $\eta < 1$ are topologically disconnected, one being a metric on elliptic space and the other being a metric on de Sitter space, and so the divergence at $\eta = 1$ in the metric \eqref{eq:PisZeroMetric} is simply a coordinate singularity.

\subsection{General Case}
In the general case with $M > 0, P > 0$, the metric \eqref{eq:FullFieldMetric} is non-diagonal. It can be diagonalized using suitable coordinate transformations in the three different cases $\Theta < 0$, $0 < \Theta < 1$, and $\Theta > 1$.

\subsubsection{\texorpdfstring{$\Theta < 0$}{Theta < 0}}
For $\Theta$ to be negative, we require from Eq. \eqref{eq:def:theta} that $\eta^2 - \zeta^2 > 1$. Recall that $\eta$ and $\zeta$ are non-negative. We define new coordinates $(a, b)$ by
\begin{subequations} \label{eq:newfields1}
\begin{align}
\eta ={}& a \cosh \left(\frac{b}{\E}\right),
\\
\zeta ={}& a \sinh \left(\frac{b}{\E}\right),
\end{align}
\end{subequations}
where $a > 1$, $b \ge 0$. The metric \eqref{eq:FullFieldMetric} becomes
\begin{align}
d\sigma^2 ={}& \frac{a^2}{a^2-1} \left[ db^2 + \frac{\E^2}{a^2(a^2 - 1)} da^2 + \E^2 \sinh^2 \left(\frac{b}{\E}\right) d\Omega_p^2 - \E^2 \cosh^2 \left(\frac{b}{\E}\right) d\Omega_m^2 \right]. \label{eq:fieldspacemetricneg}
\end{align}
Defining $c$ by $a = \mathrm{cosec}(c/\E)$ with $0 < c < \pi \E/2$, the metric becomes
\begin{align}
d\sigma^2 ={}& \sec^2 \left(\frac{c}{\E}\right) \left( db^{2} + dc^2 + \E^2 \sinh^2 \left(\frac{b}{\E}\right) d\Omega_p^2 - \E^2 \cosh^2 \left(\frac{b}{\E}\right) d\Omega_m^2 \right). \label{eq:MetMixeda}
\end{align}


\subsubsection{\texorpdfstring{$0 < \Theta \le 1$}{0 < Theta < 1}}
In this regime, $\eta > \zeta$ as previously, but with $\eta^2 - \zeta^2 \le 1$. We use the same coordinate definitions \eqref{eq:newfields1}, but with $0 \le a < 1$ and $b \ge 0$. The metric is the same as Eq. \eqref{eq:fieldspacemetricneg}. This time, define $c = \E \mathrm{sech}^{-1} (a)$ with $0 < c < \infty$, which gives
\begin{align}
d\sigma^2 ={}& \mathrm{cosech}^2\left(\frac{c}{\E}\right) \left[- db^2 + dc^2 - \E^2 \sinh^2 \left(\frac{b}{\E}\right) d\Omega_p^2 + \E^2 \cosh^2 \left(\frac{b}{\E}\right) d\Omega_m^2 \right] \label{eq:MetMixedb}
\end{align}
as the metric.

\subsubsection{\texorpdfstring{$1 \le \Theta$}{1 < Theta}}
In this region of field space, $\zeta \ge \eta$. We define coordinates $(a, b)$ by
\begin{align}
\eta ={}& a \sinh \left(\frac{b}{\E}\right),
\\
\zeta ={}& a \cosh \left(\frac{b}{\E}\right),
\end{align}
with domains of $a \ge 0$, $b \ge 0$. The metric \eqref{eq:FullFieldMetric} in these coordinates is
\begin{align}
d\sigma^2 ={}& \frac{a^2}{1+a^2} \left[ - \frac{\E^2}{a^2 (1 + a^2)} da^2 + db^2 - \E^2 \cosh^2\left(\frac{b}{\E}\right) d\Omega_p^2 + \E^2 \sinh^2\left(\frac{b}{\E}\right) d\Omega_m^2 \right].
\end{align}
If we define $c = \E \mathrm{cosech}^{-1} (a)$ with $0 < c < \infty$, the metric becomes
\begin{align}
d\sigma^2 ={}& \mathrm{sech}^2\left(\frac{c}{\E}\right) \left[- db^2 + dc^2 - \E^2 \sinh^2 \left(\frac{b}{\E}\right) d\Omega_p^2 + \E^2 \cosh^2 \left(\frac{b}{\E}\right) d\Omega_m^2 \right]. \label{eq:MetMixedc}
\end{align}


The two cases $0 < \Theta \le 1$ and $\Theta \ge 1$ are two coordinate patches on the same manifold. We see that the apparent divergence in the metric \eqref{eq:FullFieldMetric} at $\eta^2 - \zeta^2 = 1$ is just a coordinate divergence; it delineates the boundary between topologically disconnected spaces ($\Theta > 0$ and $\Theta < 0$). We show in Section \ref{sec:reasonablesection} that only one of these cases is physically realistic.

\section{Physically Viable Models\label{sec:reasonablesection}}
In this section, we impose the constraint that all kinetic terms in the Einstein conformal frame have the correct signs, in order to exclude ghosts. This requires that the field space metric have positive definite signature. Of the field space configurations, only those giving rise to the metrics \eqref{eq:MetPZeroMinus} and \eqref{eq:MetMixeda} (with $M=1$) meet this condition. We investigate the constraints this imposes on the parameters of the model.

Recall that $P$ is the number of parameters in the set $\left\{ \epsilon_T \epsilon_n, n \neq T \right\}$ which are positive, and $M=N-1-P$ is the number which are negative. The metric \eqref{eq:MetPZeroMinus} occurs when $P=0$ and $M=N-1$. This requires all $\epsilon_n$ to have the same sign, except for $\epsilon_T$ which has the opposite sign. It also requires $\Theta > 0$. 

The metric \eqref{eq:MetMixeda} occurs with the correct signature when $M=1$ and $P=N-2$. This requires all $\epsilon_n$ (including $\epsilon_T$) to have the same sign except for one (not $\epsilon_T$), which has the opposite sign. This metric also requires $\Theta < 0$.

Combining these two cases, we see that all $\epsilon_n$ (including $\epsilon_T$) must have the same sign except one, which must be opposite. We now investigate what constraints these requirements impose.

\begin{figure}[t]
    \centering
        \includegraphics[width = 0.45\textwidth]{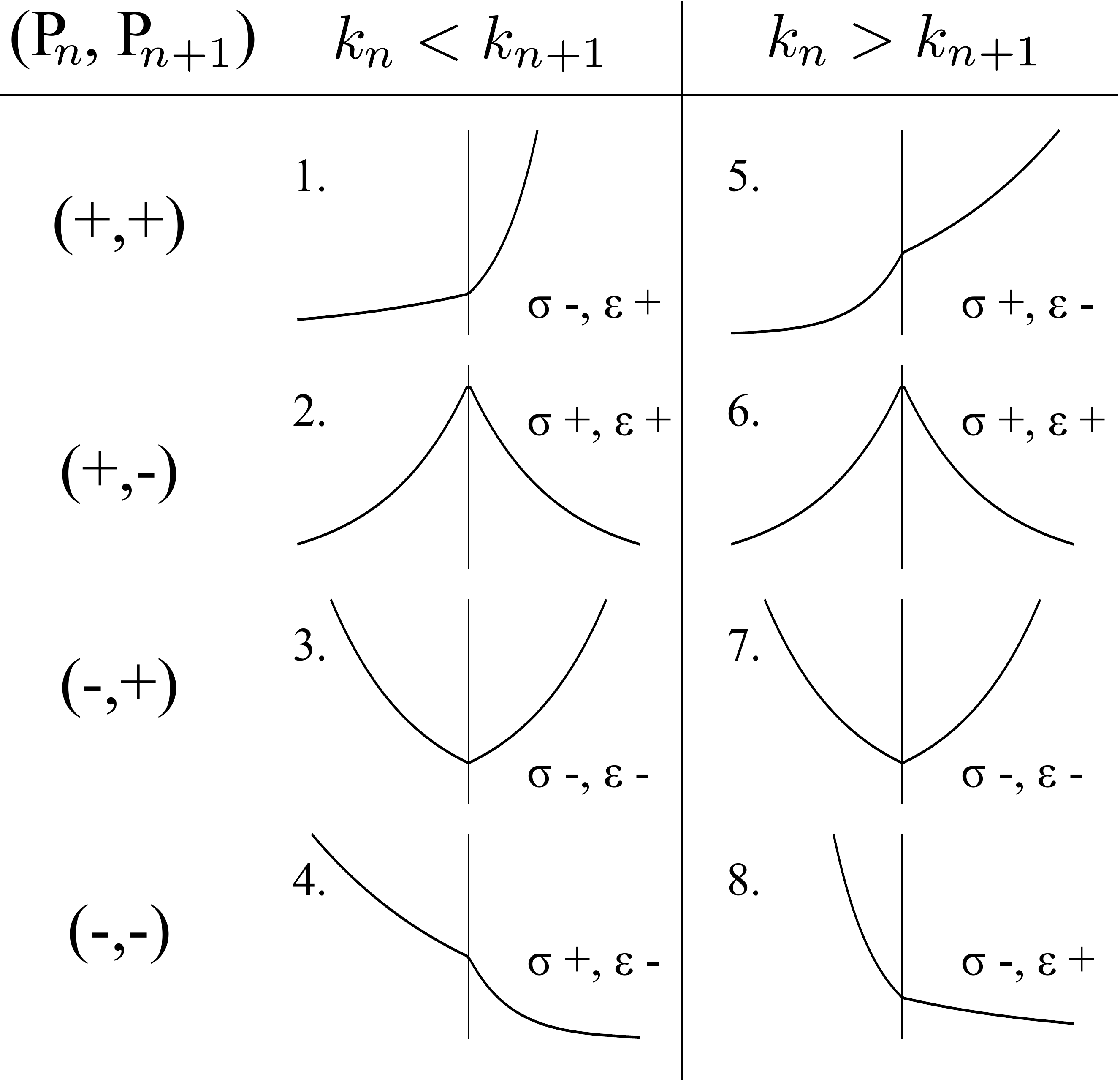}
    \caption{The behavior of the warp factor at a brane interface in the eight possible configurations. An increasing warp factor in a region has $P_n = +1$, while a decreasing warp factor has $P_n = -1$. In cases 2, 3, 6 and 7, the adjacent bulk cosmological constants can be equal. The horizontal axis in all plots is the $y$ coordinate.}
    \label{fig:Warp}
\end{figure}

At brane \bn, where the bulk regions $n$ and $n+1$ meet, there are four possible combinations for the parameters $P_n$ and $P_{n+1}$, namely $(P_n, P_{n+1}) = (-,-), (-,+), (+,-)$ and $(+,+)$. Furthermore, the bulk cosmological constant can either increase or decrease across the brane. The sign of the brane tension \lrn\sigma\ and the sign of $\epsilon_n$ for each of these eight cases is given in Fig. \ref{fig:Warp}, where the warp factor is plotted for each situation. Below, we refer to these eight possibilities as cases 1 through 8. We begin by looking at the situation where a single $\epsilon_n$ is positive ($0 \le n \le N-1$), and then look at the situation where a single $\epsilon_n$ is negative.

\subsection{\texorpdfstring{A single brane with $\epsilon_n$ positive}{A single brane with epsilon positive}}
Recall that $P_n$ is the sign of the slope of the warp factor in ${\cal R}_n$. Using $P_0 = +1$ and $P_N = -1$ (which was assumed in deriving the four-dimensional low-energy action), we need a turning point in the warp factor somewhere in the progression of branes, which restricts us to either case 2 or case 6. Both of these cases have positive $\epsilon$, and so we require that all other $\epsilon_n$ are negative. Given that if the warp factor turns back upwards after turning downwards, it would need to turn around again using another case 2 or 6 which would introduce a second positive $\epsilon$, we see that the warp factor is only allowed to increase, turn around, and then decrease. The only way to continue increasing with negative $\epsilon$ is using case 5, and the only way to decrease with negative $\epsilon$ is using case 4. Thus, the progression of cases across the branes must go
\begin{align}
5, \ldots, 5, (2 \ \mathrm{or} \ 6), 4, \ldots, 4. \label{eq:branecombo}
\end{align}
It is unnecessary to have any branes with case 5 or 4 (the first or last case may be 2/6). Note that cases 2, 4, 5 and 6 all correspond to positive tension branes.

Given the growth and fall of the warp factor, there can only be one brane on which the warp factor is a maximum. We call this the ``central'' brane. Choose $T$ to be this brane, such that $\chi(x^a, T) = 0$, and so the warp factor is unity on the brane where the warp factor is a maximum. With the progression \eqref{eq:branecombo}, $\epsilon_T = + 1$, and all other $\epsilon_n = -1$. We have $P=0$ and $M=N-1$, and so we require that $\Theta > 0$ using these field definitions.

We are interested in the sign of $\Theta$, to see if the requirement that $\Theta > 0$ is met for the metric \eqref{eq:MetPZeroMinus}. As $A_n > 0$, it is sufficient to know the sign of $A_T \Theta$. We have
\begin{align}
A_T \Theta ={}& A_T - \sum_{n \neq T} A_n e^{\chi_n}.
\end{align}
Now, given that the warp factor is a maximum on ${\cal B}_T$ and we know that $P_n = -1$ for $n > T$, it follows that $\chi_n > \chi_{n+1}$ for $n > T$. Similarly, we have $\chi_n < \chi_{n+1}$ for $n < T$. We now consider the expression for $A_n$ [Eq. \eqref{eq:def:A}] based on what we know about $P_n$ and $k_n$ from the progression \eqref{eq:branecombo}.
\begin{align}
A_T = {1}/{k_T} + {1}/{k_{T+1}}, \quad A_n = {1}/{k_n} - {1}/{k_{n+1}} \quad (n > T), \quad A_n = {1}/{k_{n+1}} - {1}/{k_{n}} \quad (n < T)
\end{align}
Thus, $\Theta$ may be written as
\begin{align}
A_T \Theta = \sum_{\substack{n \le T \\ n \neq 0}} \frac{1}{k_{n}} \left( e^{\chi_n} - e^{\chi_{n-1}} \right) + \frac{1}{k_{0}} e^{\chi_0} + \sum_{\substack{n \ge T \\ n \neq N-1}} \frac{1}{k_{n+1}} \left( e^{\chi_n} - e^{\chi_{n+1}} \right) + \frac{1}{k_{N}} e^{\chi_{N-1}}.
\end{align}
Each term in both sums is positive, and so $\Theta > 0$.

Thus, we see that a situation with all $\epsilon_n$ parameters negative bar one produces an action with no incorrectly signed kinetic terms. Furthermore, this choice of parameters requires all the brane tensions to be positive. Finally, the Ricci scalar in the action has positive coefficient, as $\epsilon_T \; \mathrm{sgn}(\Theta) = +1.$ We investigate the properties of models in this parameter space in the remainder of this paper.

\subsection{\texorpdfstring{A single brane with $\epsilon_n$ negative}{A single brane with epsilon negative}}
Here, the number of possibilities is larger than in the previous case. By using the same logic as above, we find that the following progressions of cases are the only ways to meet the required conditions:
\begin{subequations}
\begin{align}
\mbox{Option 1:}{}& \ \ 1, \ldots, 1, 5, 1, \ldots, 1, (2 \ \mathrm{or} \ 6), 8, \ldots, 8
\\
\mbox{Option 2:}{}& \ \ 1, \ldots, 1, (2 \ \mathrm{or} \ 6), 8, \ldots, 8, 4, 8, \ldots, 8
\\
\mbox{Option 3:}{}& \ \ 1, \ldots, 1, (2 \ \mathrm{or} \ 6), 8, \ldots, 8, (3 \ \mathrm{or} \ 7), 1, \ldots, 1, (2 \ \mathrm{or} \ 6), 8, \ldots, 8
\end{align}
\end{subequations}
Each of these cases requires one or more negative tension branes. We consider each of these cases in turn.

\noindent\textbf{Option 1:}\\
Let the one negative $\epsilon_n$ be $\epsilon_T$, corresponding to case 5. One brane will have the maximum warp factor; call this brane $X$. Note that $X \neq T$, as brane $T$, being case 5, does not have the maximum warp factor. We now have $\epsilon_T = -1$, and all other $\epsilon_n = +1$, and so we have $P=0$ once again, which requires $\Theta > 0$. Consider the sign of $A_T \Theta$. We have
\begin{align}
A_T \Theta ={}& A_T - \sum_{n \neq T} A_n e^{\chi_n}.
\end{align}
We can once again calculate $A_n$ explicitly.
\begin{align}
A_T = {1}/{k_{T+1}} - {1}/{k_T}, \quad &
A_n = {1}/{k_n} - {1}/{k_{n+1}} \quad (0 \le n \le X-1, n \neq T), \nonumber
\\
A_X = {1}/{k_X} + {1}/{k_{X+1}}, \quad &
A_n = {1}/{k_{n+1}} - {1}/{k_n} \quad (n > X)
\end{align}
$A_T \Theta$ can then be expressed as
\begin{align}
A_T \Theta = - \frac{1}{k_0} e^{\chi_0} - \sum_{n = 1}^{X} \frac{1}{k_n} \left( e^{\chi_n} - e^{\chi_{n-1}} \right) - \frac{1}{k_{N}} e^{\chi_{N-1}} - \sum_{n = X}^{N-2} \frac{1}{k_{n+1}} \left(e^{\chi_n} - e^{\chi_{n+1}} \right).
\end{align}
Here, all bracketed terms are positive. Thus, $\Theta < 0$, in contradiction of the requirement that $\Theta > 0$ necessary for this situation.

\noindent\textbf{Option 2:}\\
This case proceeds in exactly the same manner as Option 1, and we again find $\Theta < 0$, in contradiction of the requirements for this situation.

\noindent\textbf{Option 3.}\\
This case is a little more complicated. Let $T$ be the one brane with negative $\epsilon$, corresponding to case 3 or 7. Two branes will have a local maximum warp factor; let them be $L$ and $R$ (to the left and right of brane $T$). Now, consider $A_T \Theta$, which we require to be positive in this situation (as we once again have $P=0$).
\begin{align}
A_T \Theta ={}& A_T - \sum_{n \neq T} A_n e^{\chi_n}.
\end{align}
This time, we have
\begin{align}
A_n ={}& \frac{1}{k_n} - \frac{1}{k_{n+1}}, \ \ 0 \le n < L, \ \ T < n < R, \qquad
A_n = \frac{1}{k_{n+1}} - \frac{1}{k_{n}}, \ \ L < n < T, \ \ R < n, \nonumber
\\
A_L ={}& \frac{1}{k_L} + \frac{1}{k_{L+1}}, \qquad
A_T = \frac{1}{k_T} + \frac{1}{k_{T+1}}, \qquad
A_R = \frac{1}{k_R} + \frac{1}{k_{R+1}}.
\end{align}
Combining these, we find
\begin{align}
A_T \Theta ={}& - \frac{e^{\chi_0}}{k_0} - \sum_{n = 1}^{L} \frac{1}{k_n} \left( e^{\chi_n} - e^{\chi_{n-1}} \right) - \sum_{n = L}^{T-1} \frac{1}{k_{n+1}} \left( e^{\chi_n} - e^{\chi_{n+1}} \right) \nonumber
\\
{}& - \sum_{n = T + 1}^{R} \frac{1}{k_n} \left(e^{\chi_n} - e^{\chi_{n-1}} \right) - \sum_{n = R}^{N-1} \frac{1}{k_{n+1}} \left( e^{\chi_n}  - e^{\chi_{n+1}} \right) - \frac{e^{\chi_{N-1}}}{k_{N}}.
\end{align}
Once again, $\Theta$ is negative, and so this configuration also creates a contradiction.

\subsection{The Effect of Negative Tension Branes}
From the above arguments, we see that the only ghost-free configurations are those which do not have any negative tension branes. This is consistent with the well-known local arguments for the instability of a negative tension brane. We note that by just using positive tension branes with the assumption that $P_0 = +1$ and $P_N = -1$ (and ignoring the requirement of the different $\epsilon_n$ parameters having specific signs), the only possible combination is \eqref{eq:branecombo}, and so it is the presence of negative tension branes which are giving rise to the instability. Any valid configuration which only has positive tension branes will not have this instability.

\begin{figure}[t]
    \centering
        \includegraphics[width = 0.5\textwidth]{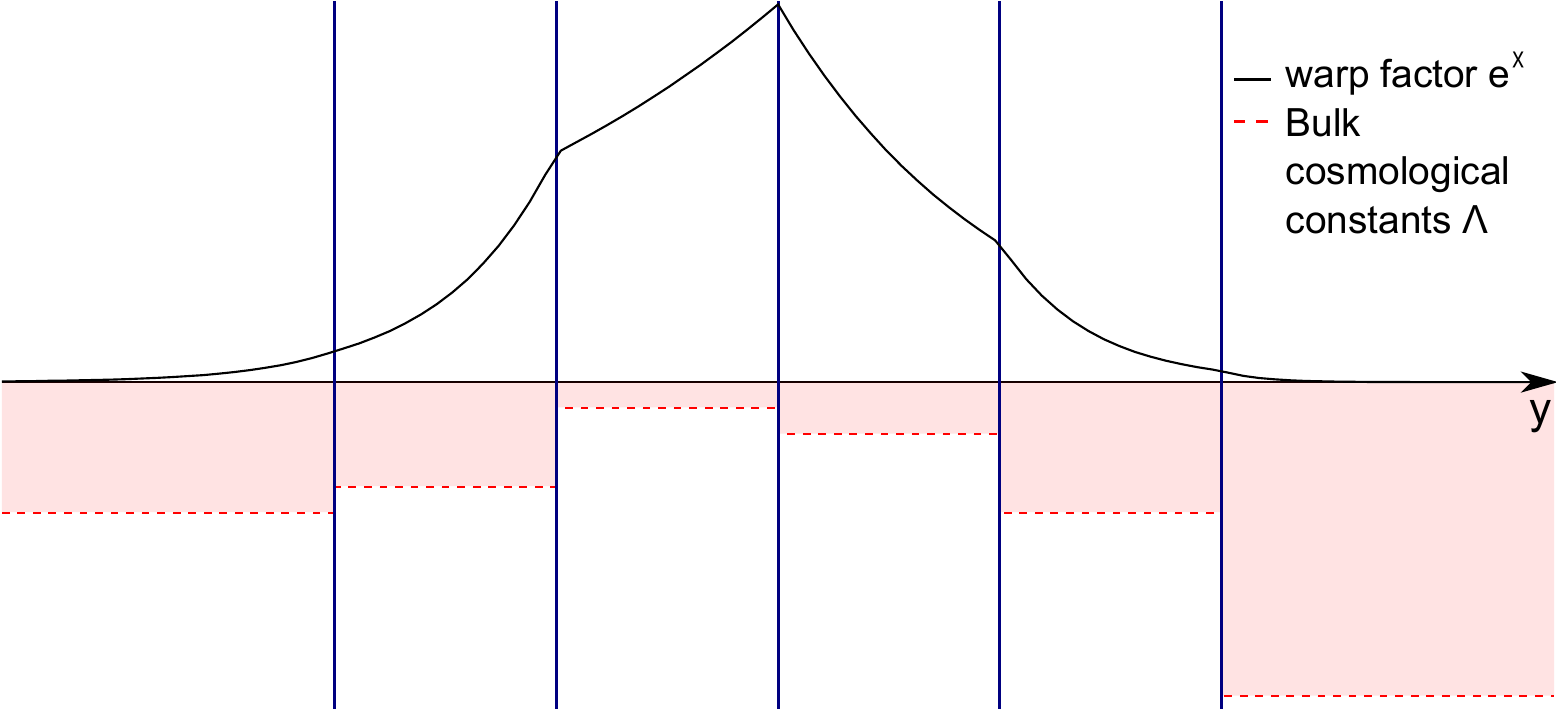}
    \caption{A diagram of the warp factor between branes and the associated bulk cosmological constants (dashed). Branes are represented as vertical lines. The bulk cosmological constants are negative, while the warp factor lies between 0 and 1.}
    \label{fig3}
\end{figure}

The combination of cases \eqref{eq:branecombo} provides a rather tight restriction on the progressions of the bulk cosmological constant which can give rise to physically viable scenarios. Recalling that the bulk cosmological constants are negative, we require the bulk cosmological constants to increase across the branes monotonically to a maximum, and then decrease monotonically (see Fig. \ref{fig3}). Note that in the special case where the first (last) brane has the maximum warp factor, then $|\Lambda|$ can be monotonically increasing (decreasing).

\section{Specializing to Physically Viable Cases\label{sec:PhysAction}}
In this section, we specialize to the physically viable cases discussed above, and find a set of variables which simplifies the action.

\subsection{The Physical Action}
We previously found that the only physically viable configuration for the model is the configuration \eqref{eq:branecombo}, in which the warp factor increases to a maximum, and then decreases again, with all brane tensions positive. We denote by $n=T$ the index of the brane with the maximum warp factor, and call this brane the ``central brane''. Specializing Eq. \eqref{eq:4DJordanOriginal} to these parameters, we find
\begin{align}
 S[\hat{\gamma}_{ab}, \psi_n, \n\phi] ={}& \int d^4 x \sqrt{-\hat{\gamma}} \frac{1}{2 \kappa_4^2} \left[ R^{(4)} \left[ \hat{\gamma}_{ab}\right] \left( 1 - \sum_{\substack{n = 0 \\ n \neq T}}^{N-1} \psi_n^2 \right) - 6 \sum_{\substack{n = 0 \\ n \neq T}}^{N-1} (\hat{\nabla}^a \psi_n) (\hat{\nabla}_a \psi_n) \right] \nonumber
\\
 & {} + \tensor[^T]S{_m}\left[\tensor{\hat{\gamma}}{_{ab}}, \tensor[^T]{\phi}{}\right] + \sum_{\substack{n = 0 \\ n \neq T}}^{N-1} \n[_m]S \left[ \frac{\psi_n^2}{B_n} \hat{\gamma}_{ab}, \n\phi \right]. \label{JordanAction}
\end{align}
This is the action in the Jordan conformal frame of the central brane.

As $P=0$, $M = N-1$, the function $\Theta$ is now given by
\begin{align}
\Theta ={}& 1 - \sum_{\substack{n = 0 \\ n \neq T}}^{N-1} \psi_n^2 = 1 - \eta^2, \label{eq:def:thetafinal}
\end{align}
and we know that $\Theta > 0$ from the arguments of the previous section. We now follow the field redefinitions \eqref{eq:def:scalarfieldsm} exactly, transforming into spherical polar coordinates. Let $(\lambda_1, \ldots, \lambda_{N-2})$ be angular coordinates such that
\begin{subequations} \label{eq:def:f}
\begin{align}
\frac{\psi_0}{\eta} ={}& \cos(\lambda_{1}) = f_0
\\
\frac{\psi_1}{\eta} ={}& \sin(\lambda_{1}) \cos(\lambda_{2}) = f_1
\\
& \vdots \nonumber
\\
\frac{\psi_{T-1}}{\eta} ={}& \sin(\lambda_{1}) \ldots \sin(\lambda_{T-1}) \cos(\lambda_{T}) = f_{T-1}
\\
\frac{\psi_{T+1}}{\eta} ={}& \sin(\lambda_{1}) \ldots \sin(\lambda_{T}) \cos(\lambda_{T+1}) = f_{T+1}
\\
& \vdots \nonumber
\\
\frac{\psi_{N-2}}{\eta} ={}& \sin(\lambda_{1}) \ldots \sin(\lambda_{N-3}) \cos(\lambda_{N-2}) = f_{N-2}
\\
\frac{\psi_{N-1}}{\eta} ={}& \sin(\lambda_{1}) \ldots \sin(\lambda_{N-3}) \sin(\lambda_{N-2}) = f_{N-1}.
\end{align}
\end{subequations}
Defining $a = \E \tanh^{-1}(\eta)$ with $a>0$ as in Section \ref{sec:PisZero}, we have our final four-dimensional low-energy action, written in the Einstein conformal frame, where $g_{ab} = \Theta \hat{\gamma}_{ab}$.
\begin{align}
  S[g_{ab}, a, \lambda_n, \n\phi] ={}& \int d^4 x \sqrt{- g} \left[ \frac{{R}^{(4)}[g]}{2 \kappa_4^2} - \frac{1}{2} \left( ({\nabla}^a a) ({\nabla}_a a) + \E^2 \sinh^2\left(\frac{a}{\E}\right) \sum_{n=1}^{N-2} \left\{ \prod_{m=1}^{n-1} \sin^2(\lambda_m) \right\} ({\nabla}^a \lambda_n) ({\nabla}_a \lambda_n) \right) \right] \nonumber
\\
 & {} + \tensor[^T]S{_m}\left[\cosh^2 \left(\frac{a}{\E}\right) \tensor{g}{_{ab}}, \tensor[^T]{\phi}{}\right] + \sum_{\substack{n = 0 \\ n \neq T}}^{N-1} \n[_m]S \left[ \sinh^2 \left(\frac{a}{\E}\right) \frac{f_n^2}{B_n^\prime} g_{ab}, \n\phi \right] \label{eq:physicalaction}
\end{align}
In a more convenient notation, the field space metric is
\begin{align}
d\sigma^2 ={}& da^2 + \E^2\sinh^2 \left(\frac{a}{\E}\right) d\Omega_n^2, \label{eq:finalfieldspacemetric}
\end{align}
where $d\Omega_n^2 = d\lambda_1^2 + \sin^2(\lambda_1) d\lambda_2^2 + \ldots$ is the metric on the unit $N-2$ sphere. This is the metric on hyperbolic space.

The target space will not be all of the quadrant of $(N-1)$-dimensional hyperbolic space for which all the field coordinates are positive, as we have yet to impose the constraint of having no branes intersecting, which was implicit in the derivation of the action. In the general case, these constraints are 
\begin{subequations}
\begin{align}
\chi_n < \chi_{n+1}, \ \ n < T,
\\
\chi_n > \chi_{n+1}, \ \ n > T,
\end{align}
\end{subequations}
where $\chi_n$ is related to $\psi_n$ by Eq. \eqref{eq:def:psi}.

\subsection{The Effect of One Brane on Another}
Given the low energy action \eqref{eq:physicalaction}, it is interesting to ask about the effect one brane has on another, depending on how they are located. To investigate this, we consider two separate scenarios, one with $N$ branes, and one with $N+1$ branes, where an extra brane has been added after the last brane in the original scenario. The effect of this extra brane on $\eta^2$ is to add an extra term to the sum \eqref{eq:def:thetafinal}. In the scenario with $N+1$ branes,
\begin{align}
\eta^{2} = \eta_0^2 + B_{N+1} e^{\chi_N},
\end{align}
where $\eta_0$ is the value of $\eta$ in the scenario with $N$ branes.

The continuity of $\chi(x^a,y)$ across branes requires that
\begin{align}
e^{\chi_N} = e^{\chi_{N-1}} e^{-2 k_N d_N},
\end{align}
where $d_N$ is the geodesic distance between the now second last and last (newly added) branes. As $\exp (\chi_{N-1}) \le 1$ ($\chi_T = 0$ is the maximum $\chi$), this contribution to $\eta^2$ becomes exponentially small as the distance to the new brane increases. Looking at Eqs. \eqref{eq:def:f}, we see that the change to the angular fields is also exponentially suppressed, and so the contribution of this new brane to the gravitational coupling is exponentially suppressed on all other branes. We therefore infer that the effect of the position of one brane on another, insofar as that information is coded into the radion fields, grows exponentially small as the distance between the branes increases. Given that the interbrane distances must be large compared to the AdS radii of curvature in order to meet the constraint from $\gamma$ (see Section \ref{sec:PPN}), this implies that the physics of a model with a large number of branes will dominated by the central brane and the nearest brane to it.

\section{Observational Constraints\label{sec:coupling}}
The theories \eqref{eq:physicalaction} that are not ruled out by instabilities contain several massless radion fields, which will mediate long range forces and give rise to corrections to general relativity. Therefore, these theories will be subject to constraints arising from Solar System and other tests of general relativity. The nature of these constraints depends on which brane normal visible matter is assumed to reside. In this section, we investigate the extent to which these radion fields modify general relativity, and determine the corresponding observational constraints on the parameters of the theory.

\subsection{Eddington PPN Parameter\label{sec:PPN}}
The Eddington parameterized post-Newtonian (PPN) parameter $\gamma$, which measures deviations from general relativity, is one of the most tightly constrained numbers from Solar System measurements of gravity. In this section, we compute this parameter from the action \eqref{eq:physicalaction}.

As shown in Ref. \cite{Flanagan2004}, for a theory of the form
\begin{align}
 S[g_{ab}, \Phi^A, \n\phi] ={}& \int d^4 x \sqrt{-g} \left\{ \frac{1}{2 \kappa_4^2} R^{(4)} [g_{ab}] - \frac{1}{2} \gamma_{AB}(\Phi^C) g^{ab} \nabla_a \Phi^A \nabla_b \Phi^B \right\} + \sum_{n=0}^{N-1} \n[_m]S \left[ e^{2 \alpha_n \left(\Phi^C\right)} g_{ab}, \n\phi \right] \label{eq:generalEinsteinframe}
\end{align}
where $\Phi^A$ are scalar fields and $\gamma_{AB}(\Phi^C)$ is the metric on field space, the Eddington PPN $\gamma$ parameter for observers on brane $n$ is given by
\begin{align}
1 - \gamma = \frac{2 \; \n[^2_0]\alpha}{1 + \n[^2_0]\alpha}
\end{align}
where
\begin{align}
\n[^2_0]\alpha = \frac{2}{\kappa_4^2}\gamma^{AB} \frac{\partial \alpha_n}{\partial \Phi^A} \frac{\partial \alpha_n}{\partial \Phi^B} \label{eq:def:alpha0}
\end{align}
and $\gamma^{AB}$ is the inverse field space metric. For our theory \eqref{eq:physicalaction}, we have $\Phi^A \equiv \left(a, \lambda_1, \ldots, \lambda_{N-2}\right)$, the field space metric is given by Eq. \eqref{eq:finalfieldspacemetric}, and the functions $\alpha_n$ are given by Eqs. \eqref{eq:def:f} and \eqref{eq:physicalaction}.

We calculate $\gamma$ for each of our branes.
On the central brane, we find that
\begin{align}
\tensor*[^T]\alpha{^2_0} = \frac{1}{3} \eta^2,
\end{align}
where $\eta = \tanh(a/\E)$ has been used. As $0 < \eta < 1$, it is possible for $\tensor*[^T]\alpha{^2_0}$ to be sufficiently small on this brane to meet experimental constraints, which require that \cite{Clifford2006}
\begin{align}
|\gamma - 1| \le 2.3 \times 10^{-5}. \label{eq:gammaconstraint}
\end{align}
This constraint implies that the brane which is closest to the central brane must be at least 5 times the bulk curvature scale away from it (from Eqs. \eqref{eq:def:psi}, \eqref{eq:def:thetafinal}, and Eq. (5.18) in Ref. \cite{Bloomfield2010}).

For the other branes, let
\begin{align} \label{eq:pn}
p(n) = \begin{cases}
n, & n < T
\\
n - 1, & n > T
\end{cases}
\end{align}
in order to account for the hole in the sum over the matter actions in Eq. \eqref{eq:physicalaction}. For brane $n$, we calculate $\tensor*[^p]\alpha{^2_0}$ to find
\begin{align}
\tensor*[^p]\alpha{^2_0} ={}& \frac{1}{3 \eta^2} \left[ 1 + (1 - \eta^2) \left\{ \sum_{j = 1}^{p} \frac{\cot^2(\lambda_j)}{\prod_{m=1}^{j-1} \sin^{2}(\lambda_m)} + (1 - \delta_{p, N-2}) \frac{\tan^2(\lambda_{p+1})}{\prod_{m=1}^{p} \sin^{2}(\lambda_m)} \right\} \right]
\\
> {}& \frac{1}{3 \eta^2}.
\end{align}
As $0 < \eta < 1$, none of these branes can give rise to a $\gamma$ parameter consistent with our observed Universe, and thus for this type of model not to be observationally excluded requires that we live on the central brane, where the warp factor is maximized. This implies that models of the form we are considering are unsuitable for explanations of the hierarchy problem, as no hierarchy can be obtained when considering Standard Model fields to be living on the central brane. Solving the hierarchy problem requires stabilizing at least some of the radion modes.

\subsection{Dark Matter Limits}
One of the motivations behind braneworld models is that the sequestering that occurs between matter on different branes may provide a natural explanation for the weakness of the coupling between normal matter and dark matter. Because of the different coupling factors of the metric to matter on different branes, there is a different Newton's constant for each brane, as well as different interaction strengths between matter on separate branes. As such, the Newton's constant becomes a Newton's matrix. In this section, we calculate the Newton's matrix measured by observers on different branes.

The Newton's matrix depends on the brane on which the observer resides, since the units in terms of which the Newton's constant is measured vary from one brane to another. As the above section constrains normal matter to live on the central brane, we calculate the Newton's matrix from the perspective of the central brane. Generalizing the arguments presented in the appendix of \cite{Bean2008b}, for a theory of the form \eqref{eq:generalEinsteinframe}, we calculate the elements of the Newton's matrix to be
\begin{align}
G_{\mathrm{eff}}^{mn} = \frac{\kappa_4^2}{8\pi} e^{2 \alpha_T} \left( 1 + \frac{2}{\kappa_5^2} \gamma^{AB} \frac{\partial \alpha_m}{\Phi^A} \frac{\partial \alpha_n}{\Phi^B} \right), \label{eq:effectiveGmn}
\end{align}
where $G_{\mathrm{eff}}^{mn}$ measures the strength of the gravitational interaction between matter on brane $n$ and matter on brane $m$. Note that for $m = n$, the quantity in the brackets is $1 + \n[^2_0]\alpha$. 

When calculating the elements of \eqref{eq:effectiveGmn}, it is again convenient to write the quantities in terms of $\eta = \tanh(a/\mu)$. We also use $p(n)$ \eqref{eq:pn}, and similarly define $q(m)$, in order to account for the missing term in the matter action sum in Eq. \eqref{eq:physicalaction}. We find
\begin{subequations} \label{eq:newtonsmatrix}
  \begin{align}
    G_{\mathrm{eff}}^{TT} ={}& \frac{\kappa_4^2}{8\pi} e^{2 \alpha_T} \left( 1 + \frac{\eta^2}{3} \right),
    \\
    G_{\mathrm{eff}}^{Tp} ={}& \frac{\kappa_4^2}{8\pi} e^{2 \alpha_T} \left( 1 + \frac{1}{3} \right),
    \\
    G_{\mathrm{eff}}^{pp} ={}& \frac{\kappa_4^2}{8\pi} e^{2 \alpha_T} \left( 1 + \frac{1}{3 \eta^2} \left[1 + (1 - \eta^2) \left\{ \sum_{j = 1}^{p} \frac{\cot^2(\lambda_j)}{\prod_{k=1}^{j-1} \sin^{2}(\lambda_k)} + (1 - \delta_{p, N-2}) \frac{\tan^2(\lambda_{p+1})}{\prod_{k=1}^{p} \sin^{2}(\lambda_k)} \right\} \right] \right),
    \\
    G_{\mathrm{eff}}^{pq} ={}& \frac{\kappa_4^2}{8\pi} e^{2 \alpha_T} \left( 1 + \frac{1}{3 \eta^2} \left[1 + (1 - \eta^2) \left\{ \sum_{j = 1}^{q} \frac{\cot^2(\lambda_j)}{\prod_{k=1}^{j-1} \sin^{2}(\lambda_k)} - \frac{1}{\prod_{k=1}^{q} \sin^{2}(\lambda_k)} \right\} \right] \right),
  \end{align}
\end{subequations}
where $m \neq n \neq T$, and $m < n$. In all cases, the ``1'' in the outermost brackets arises from graviton exchange, while the remaining terms come from the exchange of scalar quanta.

By considering the formation of the Sagittarius tidal streams, Kesden and Kamionkowski \cite{Kesden2006} have placed limits on the relative strengths of gravitational interaction between dark matter and normal matter. The constraint is roughly
\begin{align}
\left| \frac{G_{M-DM}}{\sqrt{G_{M-M} G_{DM-DM}}} - 1 \right| \lesssim 0.02 \label{eq:darkmatterconstraints}
\end{align}
where ``M'' indicates matter, and ``DM'' indicates dark matter. If we assume that all the dark matter lives on branes other than the central brane, we can calculate the constraints on our model that this provides, finding that $\eta \gtrsim 0.8$. This disagrees with the constraint \eqref{eq:gammaconstraint}, which implies $\eta \lesssim 6 \times 10^{-3}$. Thus, this model is unable to explain dark matter by positing the existence of matter fields on other branes\footnote{Note, however, that if the radion fields are stabilized, then it is possible to circumvent this restriction. As such, we can only rule out braneworld models with no moduli stabilization as an explanation for the observed weak interaction strength between dark matter and normal matter.}.

We next consider the possibility that some fraction of the dark matter lives on our (central) brane, and some fraction lives on other branes. We can then calculate the percentage of dark matter which must reside on the central brane in order to be compatible with the observational constraints \eqref{eq:gammaconstraint} and \eqref{eq:darkmatterconstraints}. On average, a mass $M$ of dark matter will be composed of a mass $\alpha M$ on our brane, say, and $(1-\alpha)M$ on other branes. The effective matter to dark matter coupling strengths will then be
\begin{subequations} \label{eq:simplenewtonsmatrix}
\begin{align}
G^{MM}_{\mathrm{eff}} ={}& G^{TT}_{\mathrm{eff}}
\\
G^{DD}_{\mathrm{eff}} ={}& G^{TT}_{\mathrm{eff}} \alpha^2 + G^{nn}_{\mathrm{eff}} (1-\alpha)^2 + G^{Tn}_{\mathrm{eff}} \alpha(1-\alpha)
\\
G^{MD}_{\mathrm{eff}} ={}& G^{TT}_{\mathrm{eff}} \alpha + G^{Tn}_{\mathrm{eff}} (1-\alpha).
\end{align}
\end{subequations}
For simplicity, we use
\begin{align}
G_{\mathrm{eff}}^{nn} ={}& G_{\mathrm{eff}}^{mn} \sim \frac{\kappa_4^2}{8\pi} e^{2 \alpha_T} \left( 1 + \frac{1}{3 \eta^2} \right)
\end{align}
as the ``off-brane to off-brane'' coupling strength. Combining values for $G^{TT}_{\mathrm{eff}}, G^{Tn}_{\mathrm{eff}}$ and $G^{nn}_{\mathrm{eff}}$ with Eqs. \eqref{eq:simplenewtonsmatrix} in the constraint \eqref{eq:darkmatterconstraints} and using $\eta^2 \sim 3.5 \times 10^{-5}$, we find $\alpha \gtrsim 0.998$, indicating that the vast majority of the dark matter must reside on our brane in this simplified model.

\section{Summary and Conclusions}

We have analyzed the observational constraints on a general five-dimensional braneworld model with arbitrary numbers of branes and without a radion stabilization mechanism, in the low energy, four-dimensional regime. The parameter space of such models was restricted by excluding ghost modes, and the phenomenology of the resulting models was analyzed. For such models to be viable, there is only one brane upon which Standard Model fields may reside, and such a configuration was unable to provide any benefit for the hierarchy problem, nor a natural explanation for the weakness of the coupling between normal matter and dark matter by sequestration. The Kaluza-Klein modes in such a model behave very similarly to the original RS-II model. Our model was not found to be ruled out experimentally, although observational constraints on the change in the value of $G_N$ between nucleosynthesis and today may do so.

Overall, we found that models with $N$ branes are quantitatively very similar to the two-brane case. Furthermore, uncompactified and orbifolded models were also found to be very similar, giving rise to identical four-dimensional low-energy theories.

\begin{acknowledgements}
We would like to thank S.-H. Henry Tye and Ira Wasserman for helpful discussions. This research was supported in part by NSF grants 0757735 and 0555216, and NASA grant NNX08AH27G.
\end{acknowledgements}

\bibliography{multibrane2}
\bibliographystyle{utphys}

\appendixpage
\appendix
\section{Results on an Orbifold\label{app:orbifold}}
In this appendix, we derive the four-dimensional low-energy action of an orbifolded $N$-brane model, and show that it is equivalent to the uncompactified model up to the rescaling of parameters.

\begin{figure}[h]
    \centering
        \includegraphics[width = 0.5\textwidth]{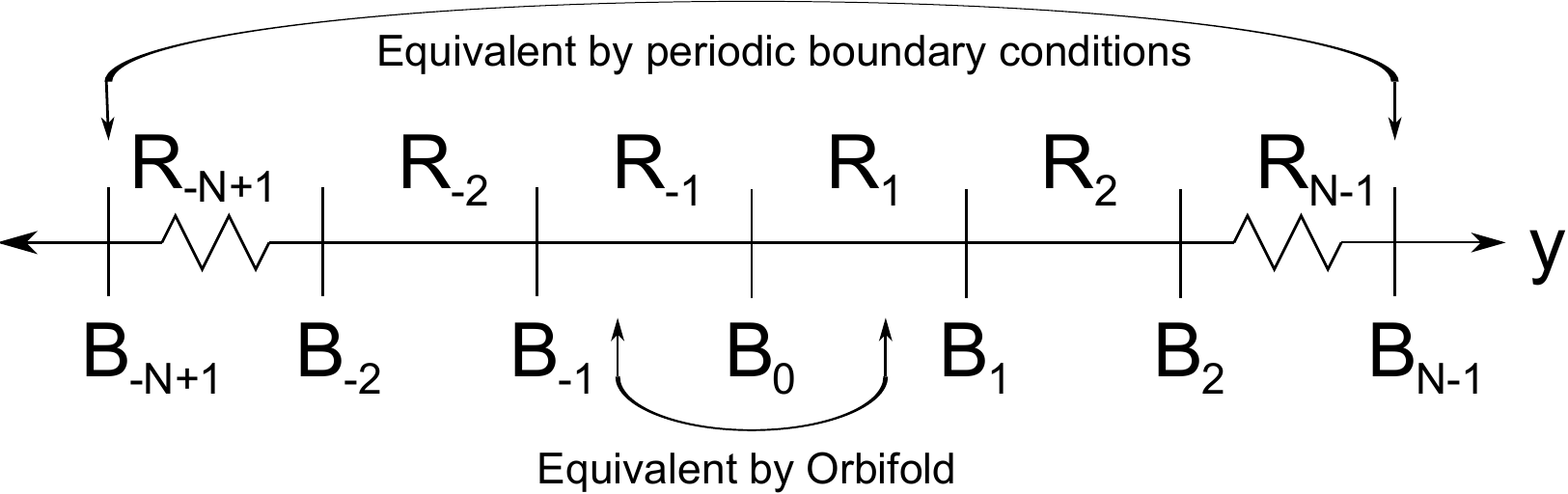}
    \caption{Diagram indicating how the branes are labeled in the construction of the orbifolded model. The orbifold symmetry identifies $y$ with $-y$, and we impose the periodicity condition of identifying $y$ with $y + 2N-2$. To calculate the action, the model is broken up into $2(N-1)$ bulk regions ${\cal R}_n$, but regions ${\cal R}_n$ and ${\cal R}_{-n}$ coincide by the orbifold symmetry.}
    \label{fig4}
\end{figure}

We begin by describing the construction of the model. We use the notation established in Ref. \cite{Bloomfield2010}. Consider a model with $N$ branes on an orbifold. The first and last branes are taken to be at the orbifold fixed points. The other $N-2$ branes lie between these two branes on one half of the orbifold, and are duplicated on the other half by the symmetry. These regions lie on a circle, and so the coordinate describing the extra dimension will be periodic. To calculate the action for this model, we take there to be $2(N-1)$ regions and $2(N-1)$ branes. Let the first brane be labeled by ${\cal B}_0$, situated at $y=0$, where $y$ is the coordinate describing the extra dimension. After gauge specializing, let there be $N-1$ branes located at $y=1, 2, \ldots, N-1$. In between the branes, we have $N-1$ bulk regions. To account for the orbifolding, continue the extra dimension in the negative $y$ direction, with another $N-1$ branes located at $y=-1, -2, \ldots, -N+1$, with the coordinates $y$ and $-y$ identified. The $y$ coordinate varies from $-N+1$ to $N-1$, and these endpoints are identified under periodic boundary conditions in $y$. The branes labeled $N-1$ and $-N+1$ are thus the same brane. Labellings are described in Fig. \ref{fig4}. The action for this model is given by
\begin{align}
S \left[ \tensor{g}{_{\Gamma \Sigma}}, \tensor[^n]{x}{^\Gamma}, \n\phi \right] ={}& \left(\sum_{n = 1}^{N-1} + \sum_{n = -N+1}^{-1} \right) \int_{\rn} d^5 \lrn{x} \sqrt{-\n g}
\left(\frac{\n[^{(5)}]R}{2 \kapfs} - \lrn \Lambda \right) + \sum_{n = -N+1}^{N-2} \frac{1}{\kapfs} \int_{\bn} d^4 \lrn{w} \sqrt{-\n h} \left(\n[^+]K + \n[^-]K \right) \nonumber
\\
& {}- \sum_{n = -N+1}^{N-2} \lrn{\sigma} \int_{\bn} d^4 \lrn{w} \sqrt{-\n h} + \tensor[^0]{S}{_m}[\tensor[^0]{h}{_{ab}}, \tensor[^0]{\phi}{}] + \frac{1}{2} \left(\sum_{n = 1}^{N-1} + \sum_{n = -N+1}^{-1} \right) \n[_m]S[\n[_{ab}]{h}, \n\phi] \label{eq:app:action}
\end{align}
The sums over branes which only run to $N-2$ are written so because the branes $-N+1$ and $N-1$ are the same brane. Note that the brane tensions at the orbifold fixed points are included once only, while the brane tensions on the other branes are doubly included. This is just a choice of how to describe the brane tensions in the orbifold. The choice of the factor of $1/2$ in the matter actions accounts for the doubling that occurs with the orbifolding.

The procedure described in Ref. \cite{Bloomfield2010} may now be followed for each region. We gauge specialize to the straight gauge, before separating length-scales in the action. Writing the metric in each region as
\begin{align}
\n[^2]{ds} = e^{\n\chi (\tensor*{x}{^c_n}, y_n)} \,\n[_{a b}]{\hat{\gamma}} (\tensor*{x}{^c_n}, y_n) dx_n^a dx_n^b + \n[^2]{\Phi} (\tensor*{x}{^c_n}, y_n) \tensor*{dy}{_n^2}
\end{align}
with $\det(\hat{\gamma}) = -1$, we can find the equations of motion at lowest order in the separation of length-scales. The following equations and boundary conditions arise, corresponding to Eqs. (3.12, 3.13, 5.3-5.7) in Ref. \cite{Bloomfield2010}. Note that the equations in regions $n$ and $-n$ are identical, as required by the orbifolding condition:
\begin{align}
\n\chi (\tensor*{w}{^c_n}, n) ={}& \np\chi (\tensor*{w}{^c_n}, n) \label{eq:app:cond1},
\\
\frac{2}{3} \kapfs \lrn{\sigma} ={}& \left.\frac{\n[_{,y}]\chi}{\n\Phi}\right|_{\lrn y = n} - \left.\frac{\np[_{,y}]\chi}{\np\Phi}\right|_{\lrnp y = n} \label{eq:app:cond2},
\\
\n[_{a b}]{\hat{\gamma}} (\tensor*{w}{^c_n}, n) ={}& \np[_{a b}]{\hat{\gamma}} (\tensor*{w}{^c_n}, n) \label{eq:app:cond3},
\\
\frac{1}{\n\Phi} \n[_{ab, y}]{\hat{\gamma}} (\tensor*{w}{^c_n}, n) ={}& \frac{1}{\np\Phi} \np[_{ab, y}]{\hat{\gamma}} (\tensor*{w}{^c_n}, n) \label{eq:app:cond4},
\\
0 ={}& \frac{1}{4} \n[^{ab}]{\hat{\gamma}} \; \n[_{bc, y}]{\hat{\gamma}} \n[^{cd}]{\hat{\gamma}} \; \n[_{da, y}]{\hat{\gamma}} - 3 \; \n[_{, y}^2]\chi - 2 \kapfs \n[^2]\Phi \lrn{\Lambda},
\\
\n[_{ad, yy}]{\hat{\gamma}} ={}& \n[_{ab,y}]{\hat{\gamma}} \; \n[^{bc}]{\hat{\gamma}} \; \n[_{cd, y}]{\hat{\gamma}} - \n[_{ad, y}]{\hat{\gamma}} \left( 2 \; \n[_{,y}]{\chi} - \frac{\n[_{,y}]\Phi}{\n\Phi}\right) \label{eq:app:solvefirst},
\\
0 ={}& \frac{1}{12} \n[^{ab}]{\hat{\gamma}} \; \n[_{bc, y}]{\hat{\gamma}} \n[^{cd}]{\hat{\gamma}} \; \n[_{da, y}]{\hat{\gamma}} + \n[_{, y}^2]\chi + \n[_{, yy}]\chi - \frac{\n[_{,y}]\Phi}{\n\Phi} \n[_{,y}]\chi + \frac{2}{3} \kapfs \n[^2]\Phi \lrn{\Lambda}.
\end{align}
The boundary conditions at the first and last branes are
\begin{align}
0 ={}& \left.\tensor[^1]{\gamma}{_{ab,y}}\right|_{y_1 = 0^+} \label{eq:app:cond5},
\\
0 ={}& \left.\tensor[^{N-1}]{\gamma}{_{ab,y}}\right|_{y_{N-1} = (N-1)^-} \label{eq:app:cond6},
\\
- P_1 \frac{1}{3} \kapfs \sigma_0 ={}& \left.\frac{\tensor[^1]\chi{_{,y}}}{\tensor[^1]{\Phi}{}}\right|_{y_1 = 0^+},
\end{align}
and
\begin{align}
P_{N-1} \frac{1}{3} \kapfs \sigma_{N-1} ={}& \left.\frac{\tensor[^{N-1}]\chi{_{,y}}}{\tensor[^{N-1}]{\Phi}{}}\right|_{y_{N-1} = (N-1)^-}.
\end{align}

Equation \eqref{eq:app:solvefirst} should be solved first. The solution (in matrix notation and suppressing indices $n$) is
\begin{align}
\boldsymbol{\hat{\gamma}} (x^a, y) ={}& \mathbf{A}(x^a) \exp\left( \mathbf{B}(x^a) \int^y \Phi(x^a, y^\prime) e^{-2\chi(x^a, y^\prime)} dy^\prime \right)
\end{align}
where $\mathbf{A}(x^a)$ and $\mathbf{B}(x^a)$ are arbitrary $4 \times 4$ matrices such that $\boldsymbol{\hat{\gamma}}$ has the properties of a metric. Combining this with Eqs. \eqref{eq:app:cond3} and \eqref{eq:app:cond4}, we see that $\mathbf{B}$ is independent of region. The boundary conditions Eqs. \eqref{eq:app:cond5} and \eqref{eq:app:cond6} then imply that $\mathbf{B} = 0$ in all regions. Finally, the condition \eqref{eq:app:cond3} then implies that $\mathbf{A}$ is independent of region, and so we can write $\n[_{ab}]{\hat{\gamma}} (x^c, y) = \hat{\gamma}_{ab} (x^c)$ for all regions.

The remaining equations of motion are then solved straightforwardly. Defining
\begin{align}
k_n = \sqrt{\frac{-\kapfs \lrn \Lambda}{6}},
\end{align}
we find
\begin{align}
\n[_{,y}]\chi = 2 P_n k_n \n\Phi
\end{align}
and the brane tuning condition
\begin{equation}
\lrn{k} \lrn{P} - \lrnp{k} \lrnp{P} = \frac{1}{3} \kapfs \lrn \sigma.
\end{equation}
For the first and last branes, this condition is
\begin{align}
k_1 P_1 ={}& -\frac{1}{6} \kapfs \sigma_0,
\\
k_{N-1} P_{N-1} ={}& \frac{1}{6} \kapfs \sigma_{N-1}.
\end{align}
The metric in each bulk region is
\begin{align}
\n[^2]{ds} = e^{\n \chi(x^c, y)} \hat{\gamma}_{ab} (x^c) dx^a dx^b + \frac{\n[^2_{,y}]\chi (x^c, y)}{4 k_n^2} dy^2.
\end{align}

Following the prescription of \cite{Bloomfield2010}, we now substitute this into the action \eqref{eq:app:action} and integrate over the fifth dimension. The result is
\begin{align}
S \left[\hat{\gamma}_{ab}, \lrn \Psi, \n\phi\right] = {}& \int d^4 x \sqrt{-\hat{\gamma}} \frac{1}{2 \kapfs} \left[ \sum_{n=1}^{N-1} \left( \frac{e^{\lrn\chi}}{k_n P_n} - \frac{e^{\lrnm\chi}}{k_n P_n} \right) R^{(4)} + \frac{3}{2} \sum_{n=1}^{N-1} \left( \frac{e^{\lrn\chi}}{k_n P_n} (\nabla \lrn\chi)^2 - \frac{e^{\lrnm\chi}}{k_n P_n} (\nabla \lrnm\chi)^2 \right) \right] \nonumber
\\
+ {}& \sum_{n=0}^{N-1} \n[_m]S[e^{\lrn\chi} \hat{\gamma}_{ab}, \n\phi],
\end{align}
where $\chi_n (x^a) = \n\chi(x^a, n)$.

We now make the following definitions.
\begin{align}
A_n ={}& \left| \frac{1}{k_n P_n} - \frac{1}{k_{n+1} P_{n+1}} \right|
\\
A_0 ={}& \left| - \frac{1}{k_1 P_1} \right| = \frac{1}{k_1}
\\
A_{N-1} ={}& \left| \frac{1}{k_{N-1} P_{N-1}} \right| = \frac{1}{k_{N-1}}
\\
\epsilon_n ={}& \mathrm{sgn} \left( \frac{1}{k_n P_n} - \frac{1}{k_{n+1} P_{n+1}} \right)
\\
\epsilon_0 ={}& \mathrm{sgn} (-P_1) = -P_1
\\
\epsilon_{N-1} ={}& \mathrm{sgn} (P_{N-1}) = P_{N-1}
\\
\Psi_n ={}& \sqrt{A_n e^{\chi_n}}
\end{align}
With these definitions, the action is given by
\begin{align}
S \left[\hat{\gamma}_{ab}, \lrn \Psi, \n\phi\right] = \int d^4 x \sqrt{-\hat{\gamma}} \frac{1}{2 \kappa_5^2} \left[ R^{(4)} \left[ \hat{\gamma}_{ab} \right] \left( \sum_{n = 0}^{N-1} \epsilon_n \Psi_n^2 \right) + 6 \sum_{n = 0}^{N-1} \epsilon_n (\nabla^a \Psi_n) (\nabla_a \Psi_n) \right] + \sum_{n = 0}^{N-1} \n[_m]S \left[ \frac{\Psi_n^2}{A_n} \hat{\gamma}_{ab}, \n\phi \right]. \label{eq:app:effaction}
\end{align}
This is identical to Eq. \eqref{eq:4DJordanBeforeGauge} above except for a factor of two multiplying $1/4\kapfs$, which arises from integrating each region twice rather than once. Otherwise, only the definitions of $\epsilon_0$, $A_0$, $\epsilon_{N-1}$ and $A_{N-1}$ have changed, which corrects for the removal of the regions between the first and last branes and infinity in the bulk. Thus, the four-dimensional low energy action for this model is the same as for the uncompactified case \eqref{eq:actionEinstein}, although some parameters have been modified. A special case of the orbifolded model is the two-brane case, the Randall-Sundrum model \cite{Randall1999}. In this case, the action \eqref{eq:app:effaction} reduces to previously known four-dimensional actions \cite{Kanno2002}.

Most of the analysis for the orbifolded scenario is identical to that for the orbifolded scenario. The only place time when the orbifolded scenario requires a separate analysis is when removing ghost modes. In the orbifolded case, we again want all $\epsilon_n$ parameters to have the same sign except for one, which is opposite. Note that we now have $\epsilon_0 = \mathrm{sgn}(\sigma_0) = -P_1$ and $\epsilon_{N-1} = \mathrm{sgn}(\sigma_{N-1}) = P_{N-1}$. For the first and last branes, we may only choose whether $\epsilon$ is positive or negative, while for the intermediary branes, all of the previously discussed cases are possibilities.

For a single positive $\epsilon_n$, we need one of the following configurations:
\begin{align*}
&{}-, 5, \ldots, 5, (2 \ \mathrm{or} \ 6), 4, \ldots, 4, -,
\\
&{}+, 4, \ldots, 4, -,
\\
&{}-, 5, \ldots, 5, +.
\end{align*}
For a single negative $\epsilon_n$, the options are
\begin{align*}
&{}-, 1, \ldots, 1, +,
\\
&{}+, 8, \ldots, 8, -,
\\
&{}+, 8, \ldots, 8, (3 \ \mathrm{or} \ 7), 1, \ldots, 1, +.
\end{align*}
The analysis of each configuration proceeds exactly as in Sec. \ref{sec:reasonablesection}. We find that we must have a single positive $\epsilon_n$, with all other $\epsilon_n$ negative. This implies that all branes must be positive tension, with the possible exception of the first and last branes, which may be negative. Again, the warp factor thus rises to a maximum and then falls again. If the first brane has the maximum warp factor, it has a positive tension, and similarly for the last brane. The four-dimensional low energy action specialized to such a configuration is described by \eqref{eq:physicalaction} above.

As the constraints on the Eddington $\gamma$ factor and the dark matter limits arise only from this action, the constraints on this orbifolded model are identical to those in the uncompactified model.

In arriving at the four-dimensional low energy action \eqref{eq:app:effaction}, we make the same approximations as in Ref. \cite{Bloomfield2010}, namely that the separation of length-scales is valid everywhere between the branes. However, we don't have any issues with the separation of length-scales breaking down towards infinity in the bulk, and nor do we need to invoke global hyperbolicity to constrain the behavior of the warp factor outside the collection of branes. Furthermore, the boundary conditions imposed by the orbifolding ensures that the degree of freedom $\mathbf{B}$ is projected out. In these regards, the orbifolded analysis is more robust than the uncompactified analysis.

\section{Kaluza-Klein Modes\label{app:KKmodes}}
In this appendix, we venture away from the four-dimensional theory to investigate the Kaluza-Klein modes of our model. The methods and results here mimic the original RS-II model \cite{Randall1999a} closely.

Consider an uncompactified model with $N$ branes (with brane tensions tuned) and no matter. The solution for the five-dimensional metric will be
\begin{align}
ds^2 = e^{\chi(y)} \eta_{ab} dx^a dx^b + dy^2
\end{align}
where $\chi_{,y} = 2 k_n P_n$, and $\chi$ is continuous. Now consider metric fluctuations of the form
\begin{align}
ds^2 = \left( e^{\chi(y)} \eta_{ab} + h_{ab}(x^c, y)\right)dx^a dx^b + dy^2.
\end{align}
Decomposing $h_{ab}$ into Fourier modes $h_{ab}(x^c, y) = h_{ab} \psi(y) \exp(i p_c x^c)$, where $p^c$ is a four-momentum with $p^2 = - m^2$, we find to first order in $h$
\begin{align}
\left( -\frac{1}{2} m^2 e^{-\chi} - \frac{1}{2} \frac{\partial^2}{\partial y^2} + \frac{1}{2} \left(\chi_{,y}\right)^2 + \frac{\chi_{,yy}}{2} \right) \psi = 0. \label{eq:KKmodes}
\end{align}
Our gauge choice is $h^a_a = \partial^a h_{ab} = 0$. Eq. \eqref{eq:KKmodes} is equivalent to Eq. (8) in \cite{Randall1999a}. As discussed there, the solutions to this equation are Bessel functions (although here, they must be defined piecewise because of the piecewise nature of $\chi$). There is a massless graviton mode, which has been integrated to give the four-dimensional effective graviton in our low-energy theory \eqref{eq:physicalaction}, and a continuum of massive Kaluza-Klein graviton modes, which in this paper were previously truncated.

As in the RS-II model, there is no mass gap. Note that there are no so-called ``ultra-light'' \cite{Kogan2000, Kogan2001, Damour2002} modes present in this model, as such modes occur in a model where the mass spectrum is quantized. Although the presence of extra branes complicates the mathematics, the physical effect of the Kaluza-Klein modes in our model is essentially the same as in the RS-II model.

In an orbifolded model, the analysis of the Kaluza-Klein modes follows similarly, but the orbifolding condition implies that the mass spectrum is quantized, and we expect ultra-light modes to be present (see \cite{Damour2002} and citations therein).

\end{document}